\documentclass[12pt]{iopart}
\usepackage{graphicx}% Include figure files
\usepackage{dcolumn}% Align table columns on decimal point
\usepackage{bm}% bold math
\usepackage{rotating}

\newcommand{\be}{\begin{equation}}
\newcommand{\ee}{\end{equation}}
\newcommand{\bea}{\begin{eqnarray}}
\newcommand{\eea}{\end{eqnarray}}
\newcommand{\bitz}{\begin{itemize}}
\newcommand{\eitz}{\end{itemize}}
\newcommand{\ben}{\begin{enumerate}}
\newcommand{\een}{\end{enumerate}}
\newcommand{\bdesc}{\begin{description}}
\newcommand{\edesc}{\end{description}}

\newcommand{\cn}{compound nucleus }
\newcommand{\cnn}{compound nucleus}
\newcommand{\cndash}{compound-nucleus }

\begin{document}

\title{Theoretical descriptions of compound-nuclear reactions: open problems \& challenges}
\author{Brett V. Carlson$^{1}$, Jutta E. Escher$^{2}$, and Mahir S. Hussein$^{3}$}
\address{$^{1}$ Depto. de F\'{\i}sica, Instituto Tecnol\'{o}gico da Aeron\'{a}utica,
12228-900 S\~{a}o Jos\'{e} dos Campos, SP, Brazil}
\address{$^{2}$ Lawrence Livermore National Laboratory, Livermore, CA 94550, USA}
\address{$^{3}$ Instituto de Estudos Avan\c{c}ados, Universidade de S\~{a}o Paulo, Caixa Postale 72012, 05508-970 S\~{a}o Paulo-SP, Brazil and Instituto de F\'{\i}sica, Universidade de S\~{a}o Paulo, Caixa Postale 66318, 05314-970 S\~{a}o Paulo-SP, Brazil}
%\pacs{}

\date{\today}% It is always \today, today,

\begin{abstract}
Compound-nuclear processes play an important role for nuclear physics applications and are crucial for our understanding of the nuclear many-body problem. 
Despite intensive interest in this area, some of the available theoretical developments have not yet been fully tested and implemented. 
We revisit the general theory of compound-nuclear reactions, discuss descriptions of pre-equilibrium reactions, and consider extensions that are needed in order to get cross section information from indirect measurements.
\end{abstract}
%\maketitle

%----- Sections -------------------------------

%--Section: INTRODUCTION ----
\section{Introduction}
\label{sec_intro}

Compound-nuclear reactions play an important role in basic and applied nuclear physics. They provide a prime example of chaotic behavior of a quantum-mechanical many-body system~\cite{Mitchell:10, Weidenmuller:09} and their cross sections are required for nuclear astrophysics, national security, and nuclear-energy applications.
The theoretical formalisms used to describe compound reactions are typically considered well-established. R-matrix treatments~\cite{Lane:58,Descouvemont:10} are employed for reactions proceeding through isolated resonances and a combination of Hauser-Feshbach theory~\cite{HauserFeshbach:52} and pre-equilibrium descriptions is used for reactions involving strongly-overlapping resonances.
Computational tools are readily available and extensively used in the nuclear science community for calculating a wide range of cross sections. Recommendations for input models and parameters have been formulated~\cite{Capote:09shrt} and evaluations using these tools have been published.

Hauser-Feshbach calculations require inputs that are typically obtained from complementary measurements (low-lying nuclear levels, separation energies, etc.) and nuclear-structure models. In the last decade or so, there have been increased efforts to move from phenomenological models (for level densities, $\gamma$-ray strength functions, etc.) to more microscopic structure descriptions. While the microscopic descriptions may not be as successful in reproducing measured cross sections as their phenomenological counterparts at this time, they are crucial for developing a more predictive treatment of compound reactions.  

Equally important in this context is a reconsideration of the reaction models that are implemented in the codes and of the underlying assumptions and approximations. Much less attention has been paid to this aspect, but modern computational capabilities and also the improved structure models that are now available should make it possible to develop better treatments of the reaction mechanisms involved in compound reactions.
This is particularly relevant if one wants to calculate cross sections involving isotopes more than a few units away from the valley of stability.  Here, some of the assumptions underlying current descriptions may no longer be valid and extensions of the reaction models may become necessary.

For instance, as one moves away from the valley of stability, the nuclei under consideration (which may serve as targets in neutron-rich astrophysical environments) become weakly-bound. The associated low level densities make it relevant to question the statistical assumptions underlying the standard Hauser-Feshbach formalism and revisit the issue of energy averaging in compound nuclei.  Intermediate structure may become important, thus requiring an explicit treatment of doorway states. Also, experimental restrictions in radioactive-beam experiments may not allow for an experimental separation of direct and compound contributions to measured cross sections. A formalism that can treat both on the same footing would be desirable~\cite{Goriely:97, Goriely:98, Chiba:08}.

Not all cross sections of interest can be measured directly. Radioactive-beam experiments can probe reactions in inverse kinematics, but such experiments typically provide information on transfer reactions (such as (d,p)) or inelastic scattering, which have to be related to the reaction of interest (e.g. a neutron-induced reaction). Theory is needed to describe the formation of a compound nucleus (or lack thereof) in such transfer or inelastic scattering reactions and to extract the desired cross section from the indirect measurement~\cite{Escher:12rmp}.

Recent studies have begun to address some of these issues. Here we give a brief summary of existing theories for compound-nuclear, pre-equilibrium, and hybrid reactions.  In the next section, Section~\ref{sec_cnTheory}, we focus on the theory of the compound nucleus from a projection-operator perspective. Section~\ref{sec_preeq} covers the current status of pre-equilibrium theories. Section~\ref{sec_hybrids} discusses hybrid reactions - indirect methods for determining cross sections of interest. 
In Section~\ref{sec_Challenges}, we collect some of the open questions and challenges that remain to be addressed.
Concluding remarks are offered in Section~\ref{sec_summary}.

%--Section: Theory of the Compound Nucleus
\section{Theory of the Compound Nucleus}
\label{sec_cnTheory}

The idea of a fully equilibrated `compound' nuclear system, whose decay is independent of its formation, was introduced in 1937 by Niels Bohr~\cite{Bohr:36}. The equilibrium concept was used by Weisskopf and Ewing~\cite{WeisskopfEwing:40} for a calculation of particle emission from a compound nucleus (CN), and a quantum-mechanical treatment was provided by Hauser and Feshbach \cite{HauserFeshbach:52}.
Hauser-Feshbach theory is widely used today, although refinements are needed to account for various experimental findings.
For instance, Kawai, Kerman and McVoy~\cite{Kawai:73} generalized the theory to cases where direct reactions are present.
The profound statistical concepts underlying compound reactions were recognized~\cite{Mitchell:10, Weidenmuller:09} and led to the consideration of observables other than average cross sections, most notably correlation functions, first discussed by Ericson~\cite{Ericson:60,Ericson:63}.
Many of the ideas developed since the introduction of the compound-nucleus concept found their way into other branches of physics, such as condensed-matter physics and quantum chromodynamics. Here we focus on the theory of the compound nucleus from a projection-operator perspective and summarize relevant developments.

\subsection{Bohr's Hypothesis and the Hauser-Feshbach cross section}
\label{sec_cnTheory_Bohr}
The Bohr independence hypothesis states that formation and decay of the compound nucleus are independent of each other. Consequently, the cross section takes a product form (we omit the kinematic factor and consider a fixed the angular momentum value):
\begin{equation}
\sigma_{cc^{\prime}} = \xi_{c} \cdot \xi_{c^{\prime}}
\end{equation}
Unitarity requires that the sum over the final channels give the probability of the formation of the compound nucleus in channel $c$. This probability is just the transmission coefficient, $T_c$,
\begin{equation}
T_{c} = \sum_{c^{\prime}} \sigma_{cc^{\prime}} = \xi_{c} \sum_{c^{\prime}}  \xi_{c^{\prime}} \;\;\;\;\;\;\;
\mbox{     or     } \;\;\;\;\;\;\;
 \xi_{c} = \frac{T_{c}}{\sum_{c^{\prime}}  \xi_{c^{\prime}}}
 \end{equation}
Summing the above equation over $c$ gives
$\sum_{c} T_{c} = \left(\sum_{c}  \xi_{c}\right)^2$, 
which leads to 
$\xi_{c} = T_{c} / \sqrt{\sum_{c^{\prime\prime}} T_{c^{\prime\prime}}}$.
Thus we obtain the Hauser-Feshbach form of the compound nucleus cross section:
\begin{equation}
\sigma_{cc^{\prime}} = \frac{T_{c} T_{c^{\prime}}}{\sum_{c^{\prime\prime}} T_{c^{\prime\prime}}} \; .
\label{eq:HF}
\end{equation}
This expression was obtained by using the Bohr hypothesis and unitarity. In the present form, it neglects correlations between the different channels, which are known to be important, in particular at low energies. In practical applications, these correlations can be accounted for by introducing a `width fluctuation correction factor,' $W_{cc^{\prime}}$, in the expression for the cross section:
\begin{equation}
\sigma_{cc^{\prime}} = W_{cc^{\prime}} \frac{T_{c} T_{c^{\prime}}}{\sum_{c^{\prime\prime}} T_{c^{\prime\prime}}} \; ,
\end{equation}
where
\begin{equation}
W_{cc^{\prime}} \equiv  \left< \frac{T_{c} T_{c^{\prime}}}{\sum_{c^{\prime\prime}} T_{c^{\prime\prime}}} \right> \; / \;\;
\frac{\langle T_{c} \rangle \langle T_{c^{\prime}} \rangle }{\sum_{c^{\prime\prime}} \langle T_{c^{\prime\prime}} \rangle }   \; .
\label{eq:WFCF}
\end{equation}
and the brackets denote averaging over a suitable energy interval $I$.
For the case of elastic scattering ($c = c^{\prime}$),
these correlations cause an enhancement of the cross section, $2 \le W_{cc} \le 3$, as we show in Section~\ref{sec_cnTheory_KKM} below.

\subsection{Feshbach's theory of compound nuclear reactions}
\label{sec_cnTheory_Feshbach}

Feshbach \cite{Feshbach:58,Feshbach:62} introduced a formal theory of nuclear reactions based on projection operators that serve to isolate the reaction mechanisms of interest. In this approach, the complicated (closed) channels are eliminated from explicit consideration via projection and subsequent energy averaging. The formalism covers both direct and compound processes and is applicable to elastic scattering as well as more complex reaction mechanisms. Here we summarize the salient features for the elastic case.

We assume one open channel (elastic) and many closed channels (the compound nuclear states) and denote the total wave function of the system by $|\Psi>$. The open channel is projected from the wave function by using the projection operator $P$, while the closed channels are projected out with the projection operator $Q$. The Schr\"{o}dinger equation of the system, $(E - H) | \Psi> = 0$,
becomes a set of coupled equations:
\begin{equation}
(E - PHP)P|\Psi> = PHQ Q|\Psi>
\end{equation}
\begin{equation}
(E - QHQ) Q|\Psi> = QHP P|\Psi>
\end{equation}
The effective equation for $P|\Psi>$ is obtained by eliminating the $Q|\Psi>$ component,
\begin{eqnarray}
&&(E - PHP - PHQ\frac{1}{E - QHQ}QHP) P|\Psi> \nonumber \\
&&= (E- PHQG_{Q}QHP)P|\Psi> = 0  \;,
\label{eq:effPPsi}
\end{eqnarray}
where we have introduced the Q-space Green's function:
\begin{eqnarray}
G_{Q} = \frac{1}{E - QHQ}
\label{eq:QGreen}
\end{eqnarray}
Equation~(\ref{eq:effPPsi}) is very complicated to solve. It contains a strongly energy-dependent effective potential through the term $PHQ G_{Q} QHP$. There is no absorption, as the effective $P$-Hamiltonian $PHP$ is Hermitian. The standard procedure for simplifying the theory is to introduce an energy-averaged Q-space Green's function, $< G_Q >$. This energy-average can be calculated using a Lorentzian weight function of the form $\rho(E, E^{\prime}) = (I/2\pi)[(E^{\prime} - E)^2 + (I/2)^2]$, with an averaging width $I$ that is much larger than the width of a typical CN resonance, $I \gg \Gamma$. Using contour integration we obtain
\begin{equation}
< G_Q > = \frac{1}{E - QHQ + iI/2} \; .
\label{eq:EAvQGreen}
\end{equation}
When $<G_Q >$ is used instead of $G_Q$ in Eq.~(\ref{eq:effPPsi}), one speaks of the optical model equation. The effective potential now has a slow energy dependence, but is intrinsically complex. The loss of flux from the elastic channel described by the complex part of optical potential $PHQ \langle G_{Q}\rangle QHP$  owes its origin to the underlying CN coupling. This flux is not lost, as it reappears through the fluctuation cross section, as we describe below.
Equation~(\ref{eq:effPPsi}) has the following formal solution:
\begin{equation}
P|\Psi> = |\phi> + G^{(+)}_{P}  PHQ \frac{1}{E - QHQ}QHP|\Psi> \; .
\end{equation}
Here $G^{(+)}_{P}$ is the ``free'' P-space Green's function, given by $G^{(+)}_{P} = 1/ (E - PHP + i\epsilon)$.

The $S$ matrix can then be written as
\begin{equation}
S = S^{(0)} - 2\pi i PHQ \frac{1}{E - QHQ - QHP G^{(+)}_{P} PHQ}QHP \; .
\end{equation}
The first term is a unitary background term, while the second term contains a sum over resonances.
These resonances are the compound nucleus resonances. To make connection with the complex optical potential, one resorts to energy averaging over these resonances. The average cross section is found to contain three terms: the potential scattering term, the compound nucleus term and the interference term. This last term makes the analysis of reaction data rather cumbersome
and complicates attempts to separate the cross sections into direct and compound contributions.
Kawai, Kerman, and McVoy (KKM)~\cite{Kawai:73} offered a solution, which we describe next. 
 
\subsection{The Kawai-Kerman-and-McVoy  (KKM) theory}
\label{sec_cnTheory_KKM}
In Ref.~\cite{Kawai:73}, Kawai, Kerman, and McVoy (KKM) derive a new representation of the full $S$ matrix which eliminates the compound-direct interference term. They introduce the optical potential at the outset when dealing with the effective equation for the $P|\Psi>$ wave function. This procedure results in an average cross section containing the optical (or coupled-channels) term and the compound-nucleus, or ``fluctuation,'' term; the interference term averages to zero. 

More specifically, Kawai, Kerman, and McVoy start with Equation~(\ref{eq:effPPsi}) and add and subtract the energy-averaged Q-space Green's function, Eq.~(\ref{eq:EAvQGreen}), to obtain:
\begin{equation}
(E - PHP - PHQ<G_{Q}>QHP) P|\Psi> = PVQG_{Q} QVPP|\Psi>
\end{equation}
where $PVQ = PHQ\sqrt{\frac{I/2}{E - QHQ + iI/2}}$. A similar form is found for $QVP$.

Imposing strict time-reversal invariance, the scattering S-matrix for a reaction $c \rightarrow c^{\prime}$ which proceeds through compound-nucleus states, $|q>$, can be written as
\begin{equation}
S_{cc^{\prime}} = S^{opt}_{cc^{\prime}}-i \sum_{q}\frac{g_{qc}g_{qc^{\prime}}}{E - \varepsilon_q} = S^{opt}_{cc^{\prime}} + S^{fl}_{cc^{\prime}}
\end{equation}
where the resonance decay amplitude $g_{qc}$ is given by,
\begin{equation}
g_{qc} =\sqrt{2\pi}\left<\phi_q|V|\psi^{(+)}_{c}\right>,
\end{equation}
and $\varepsilon_q = E_q - i\Gamma_{q}/2$ is the complex energy of the CN resonance.
The energy-averaged piece of the S-matrix is denoted by $S^{opt}$, such that
$S^{opt}_{cc^{\prime}} = \left<S_{cc^{\prime}}\right> $,
and accordingly we have the constraint
\begin{equation}
\left<S^{fl}_{cc^{\prime}}\right> = -i \left<\sum_{q}\frac{g_{qc}g_{qc^{\prime}}}{E - E_q}\right> = 0 \; .
\end{equation}
This constraint makes it possible to write the energy-averaged cross section as an incoherent sum of two terms, namely,
\begin{equation}
\sigma_{cc^{\prime}} = \sigma^{opt}_{cc^{\prime}} + \hat{\sigma}^{fl}_{cc^{\prime}}
\end{equation}
The average partial fluctuation cross section cross section, for a given orbital angular momentum value $L$, is $\hat{\sigma}^{fl}_{cc^{\prime}}$ = $\sigma^{fl}_{cc^{\prime}}$ $\pi (2L + 1)/k_{c}^{2}$,
where
\begin{equation}
\sigma^{fl}_{cc^{\prime}} = \left<|S^{fl}_{cc^{\prime}}|^2\right> = \left<\sum_{qq^{\prime}} \frac{g_{qc}g_{qc^{\prime}}g^{\star}_{qc}g^{\star}_{qc^{\prime}}}{(E - E_q + i(\Gamma_q)/2)(E - E_{q^{\prime}} - i(\Gamma_{q^{\prime}})/2} \right>  \; .
\end{equation}
To perform the energy average, one first collapses the double sum over $q$ and $q^{\prime}$ into a single sum, and uses box averaging:
\begin{equation}
\sigma^{fl}_{cc^{\prime}} = \sum_{q} (1/I) \int_{E - I/2}^{E + I/2} dE^{\prime} \frac{g_{qc}g_{qc^{\prime}}g^{\star}_{qc}g^{\star}_{qc^{\prime}}}{(E - E_q)^2 + (\Gamma_{q}/2)^2}
\end{equation}

Since the averaging interval $I$ is much larger than the average compound nucleus width, we may assume $I/2 \gg E - E_q$, without loss of generality. Allowing the integration limits to go to $\infty$, the integral over can be performed by contour integration, giving
$
\sigma^{fl}_{cc^{\prime}} = \frac{2\pi}{I}\sum_{q}[\frac{1}{\Gamma_{q}}g_{qc}g_{qc^{\prime}}g^{\star}_{qc}g^{\star}_{qc^{\prime}}]
$.
This sum can be related to an average over $q$, by recognizing $\sum_{q}[\frac{1}{\Gamma_q}g_{qc}g_{qc^{\prime}}g^{\star}_{qc}g^{\star}_{qc^{\prime}}] = (I/D) <[\frac{1}{\Gamma_q}g_{qc}g_{qc^{\prime}}g^{\star}_{qc}g^{\star}_{qc^{\prime}}]>_{q}$, where $D$ is the average spacing between CN resonances, and the ratio, $I/D$ is the number of resonances within the averaging interval. Removing the inverse of the width from inside the $q$ average, and replacing $1/\Gamma_q$ by $1/\overline{\Gamma}_q \equiv{1/ \Gamma}$, we finally get

\begin{equation}
\sigma^{fl}_{cc^{\prime}} = \frac{2\pi}{D\Gamma}<g_{qc}g_{qc^{\prime}}g^{\star}_{qc}g^{\star}_{qc^{\prime}}>_q
\end{equation}

We use the Gaussian distribution of the form factors $g$ to reduce the above average as $<g_{qc}g_{qc^{\prime}}g^{\star}_{qc}g^{\star}_{qc^{\prime}}> = < g_{qc}g^{\star}_{qc}>_{q}< g_{qc^{\prime}} g^{\star}_{qc^{\prime}}>_{q} + <g_{qc}g^{\star}_{qc^{\prime}}>_{q} < g_{qc^{\prime}}g^{\star}_{qc}>_{q} + < g_{qc}g_{qc^{\prime}}>_{q} < g^{\star}_{qc}g^{\star}_{qc^{\prime}}>_{q}$. This decomposition suggests the following form for the fluctuation cross section:
\begin{equation}
\sigma^{fl}_{cc^{\prime}} = \left<|S^{fl}_{cc^{\prime}}|^2\right> = X_{cc}X_{c^{\prime}c^{\prime}} + X_{cc^{\prime}}X_{c^{\prime}c} + (\frac{2D}{\pi \Gamma}) |Y_{cc^{\prime}}|^2 \; ,
\end{equation}
where the $X$ and $Y$ matrices are defined as follows:
\begin{equation}
X_{cc^{\prime}} = \sqrt{\frac{2\pi}{\Gamma D}}\left<g_{qc}g^{\dagger}_{qc^{\prime}}\right>
\end{equation}
\begin{equation}
Y_{cc^{\prime}} = \frac{\pi}{D}\left<g_{qc}g_{qc^{\prime}}\right>\
\end{equation}

The cross section is usually expressed in terms of the transmission or penetration matrix whose $cc^{\prime}$ element is defined by:
\begin{eqnarray}
P_{cc^{\prime}} &=& \delta_{cc^{\prime}} - \sum_{c^{\prime\prime}}S^{opt}_{cc^{\prime\prime}}(S^{opt}_{c^{\prime\prime}c^{\prime}})^{\ast} = \sum_{c^{\prime\prime}}\left<S^{fl}_{cc^{\prime\prime}}(S^{fl}_{c^{\prime\prime}c^{\prime}})^{\ast}\right> \\
&=& \sum_{c^{\prime\prime}} \left<  \sum_{qq^{\prime}}\frac{g_{qc}g_{qc^{\prime\prime}}g^{\ast}_{q^{\prime}c^{\prime\prime}}g^{\ast}_{q^{\prime}c^{\prime}}}{(E - E_q)(E - E^{\ast}_{q^{\prime}})}  \right> \; . \nonumber
\end{eqnarray}
This can be reduced to the following expression for the matrix:
\begin{equation}
P = XTr(X) + X^2 + (\frac{2D}{\pi\Gamma}) YY^{\dagger}
\end{equation}

In the region of overlapping resonances the last term, which contains the matrix $Y$, is very small and can be neglected in both $P$ and in the expression for $\sigma^{fl}_{cc^{\prime}}$. If we further ignore the $X^2$ term in the equation for the transmission coefficient above, we can then solve for $X(P)$, as $X = P/(TrP)^2$, and thus obtain the Hauser-Feshbach cross section in the presence of direct reactions,
\begin{equation}
\sigma^{fl}_{cc^{\prime}} = \frac{P_{cc}P_{c^{\prime}c^{\prime}} + P_{cc^{\prime}}P_{c^{\prime}c} }{Tr P}
\end{equation}
For the case of elastic scattering, $c = c^{\prime}$, we obtain the cross section
$
\sigma^{fl}_{cc} = 2 \frac{(P_{cc})^2}{Tr P},
$
which clearly exhibits the elastic enhancement factor of 2. This factor was verified experimentally by Kretschmer {\em et al}~\cite{Kretschmer:78}. In the opposite limit of isolated resonances, where we have weak absorption, the $g's $ are real and the matrices $\left <g_{qc}g_{qc^{\prime}}\right> = \left<g_{qc}g^{\dagger}_{qc^{\prime}}\right>$; thus  the matrix $Y = \sqrt{2D/\pi\Gamma}X$ and the fluctuation cross section for elastic scattering becomes
$
\sigma^{fl}_{cc} = 3 \frac{(P_{cc})^2}{Tr P}
$,
i.e. it carries an elastic enhancement factor of 3. In this same weak absorption limit, the cross section for a given transition is simplified significantly, as the transmission coefficients can be related to the partial width $\Gamma_c$ and density of states $(1/D)$ as $P_{cc} = 2\pi \Gamma_c /D$ (for neutrons). The leading term of the CN cross section becomes $D \Gamma_c \Gamma_{c^{\prime}}/\Gamma$, clearly exhibiting the dependence on the CN density of states. To get a full picture of this dependence, the other terms in the cross section must also be calculated~\cite{Kawai:73}.

\subsection{Intermediate structure and Doorway resonances}
\label{sec_cnTheory_doorways}

It was recognized in the early 1960s that the average cross section exhibits modulations with a width that is much larger than the average CN resonance width. The concept of Intermediated Structure was introduced~\cite{Kerman:63}, and an interpretation in terms of a``Doorway" state was advanced \cite{Block:63} (see also \cite{Bolsterli:66}). The structure of a doorway state is considered to be much simpler than that of a CN resonance. To account for intermediate structure in reactions, one introduces a third projection operator, $D$,  in addition to the projection operators $P$ and $Q$, such that $P + Q + D = 1$ with $PQ = PD = QD = 0$ and $PP = P$, $QQ = Q$, and $DD = D$. The doorway state's width is composed of two terms, the escape width, which measures the degree of the coupling to the open channels, and the spreading width, which measures the strength of the coupling to the more complicated CN states.

The system of coupled equations previously considered is now replaced by a system consisting of three coupled equations:
\begin{equation}
(E - PHP) P|\Psi> = PHQ Q|\Psi > + PHD D|\psi>
\end{equation}
\begin{equation}
(E - QHQ)Q|\Psi> = QHP P|\Psi> + QHD D|\Psi>
\end{equation}
\begin{equation}
(E - DHD) D|\Psi> = DHP P|\Psi> + DHQ Q|\psi>
\end{equation}
The formal solution of the above set of equations follows basically the same steps as before. It is, however, more complicated, because of the additional couplings that can occur. One may resort to the extreme doorway model in which direct coupling between the $P$ and $Q$ spaces is absent and can only proceed through the doorway. This implies $PHQ = 0$ and $QHP = 0$. The three equations then become
\begin{equation}
(E - PHP) P|\Psi> = PHD D|\psi>
\end{equation}
\begin{equation}
(E - QHQ)Q|\Psi> = QHD D|\Psi>
\end{equation}
\begin{equation}
(E - DHD) D|\Psi> = DHP P|\Psi> + DHQ Q|\psi> \; .
\end{equation}
The next step in the reduction is to average over the $Q$ resonances. We first eliminate the $Q$-component of the wave function, recognizing that $P|\Psi> = |\phi> + G_{P} PHD D|\Psi>$, where $|\phi>$ is a solution of $(E - PHP)|\phi> = 0$ and $Q|\Psi> = G_{Q}QHD D|\psi>$. We introduce the energy-averaged $Q$ Green's function $<G_{Q}> = <\frac{1}{E - QHQ}> = \frac{1}{E - QHQ + i(I/2)}$. The width of the averaging interval $I$ is much larger than the width of a typical CN resonance, but smaller than than the width of the doorway, $\overline{\Gamma}_q \ll$ $I < \Gamma_D$.  Using $<G_{Q}>$ in the third equation, which is now an equation for the $Q$-space averaged doorway wave function, we obtain
\begin{equation}
\!\!\!\!\!\!\!\!\!\!\!\!\!\!\!\!\!\!\!\!\!\!\!\!\!\!\!\!\!\!\!\!\! (E - DHD - DHQ <G_{Q}>QHD - DHPG_{P}PHD) D|\Psi> = DHP |\phi>
\end{equation}

Inserting this solution back in the first equation for $P|\Psi>$ results in an equation for the open channel's wave function.
The Q-space averaged $S$-matrix is now directly obtained. Note that the doorway width is composed of an escape width, $\Gamma^{\uparrow}_D$, and a spreading (damping) widths, $\Gamma^{\downarrow}_D$, i.e. $\Gamma_D =\Gamma^{\uparrow}_D + \Gamma^{\downarrow}_D$. In the simple case of purely elastic scattering, the $S$-matrix becomes
\begin{eqnarray}
S_{00} &=& S^{(0)}_{00} \left[1 - i\frac{\Gamma^{\uparrow}_{D}}{E - E_D + i(\Gamma^{\uparrow}_{D} + \Gamma^{\downarrow}_{D})/2}\right] \nonumber \\
&=& S^{(0)}_{00}\left[\frac{E - E_D - i(\Gamma^{\uparrow}_D - \Gamma^{\downarrow}_D)/2}{E - E_D + i(\Gamma^{\uparrow}_D + \Gamma^{\downarrow}_D)/2}\right] \; ,
\end{eqnarray}
which is manifestly non-unitary owing to the presence of the spreading width.  The transmission coefficient can be calculated as $P_{00} = 1 - |S_{00}|^2$ and we find
\begin{equation}
P_{00} = \frac{\Gamma^{\uparrow}_D \Gamma^{\downarrow}_D}{(E- E_D)^2 + (\Gamma_{D}/2)^2}
\end{equation}
This expression is symmetric around the doorway energy $E_D$. The symmetry is removed when the couplings $PHQ$ and $QHP$ are included, which is necessary, e.g., for the case of Isobaric Analog Resonances~\cite{Feshbach:58,Feshbach:62}.

The concept of doorway states has been of great value in the description of nuclear reactions involving resonances. Doorway states can be as simple as a two particle-one hole state or a collective giant resonance, which is a coherent excitation of one particle-one hole states. Both are damped due to their couplings to the more complex CN states. The doorway idea is also used in other branches of physics, such as in the description of metal clusters. Furthermore, the concept of doorway states with increasing complexity is the basic ingredient in pre-equilibrium theories, such as the one developed by Feshbach, Kerman, and Koonin~\cite{Feshbach:80}. This topic will be discussed in Section~\ref{sec_preeq}.

\subsection{Eriscon's fluctuations}
\label{sec_cnTheory_ericson}

In the region of overlapping resonances, the peaks in the cross section do not correspond to individual resonances. The cross section is just ``noise"  arising from the complex underlying CN fluctuation and information about the compound nucleus, such as life time, density of states, etc., cannot be simply obtained.  In the early 1960s T. Ericson suggested measuring the cross section correlation function in order to extract this information. The correlation function is defined by

\begin{equation}
C_{c c^{\prime}}(\varepsilon) = \langle\sigma_{cc^{\prime}}(E) \sigma_{cc^{\prime}}(E + \varepsilon)\rangle
\end{equation}

The calculation of $C_{c c^{\prime}}(\varepsilon)$ is closely related to the calculation of the so-called $S$-matrix correlation function. As a matter of fact, from the definition of the cross section in terms of the $S$-matrix, we have,

\begin{equation}
\langle\sigma_{cc^{\prime}}(E) \sigma_{cc^{\prime}}(E + \varepsilon)\rangle = \langle S_{cc^\prime}(E)S^{\dagger}_{cc^{\prime}}(E)S_{cc^\prime}(E+\varepsilon)S^{\dagger}_{cc^{\prime}}(E + \varepsilon)\rangle
\end{equation}

Assuming that the $S$ matrix is a Gaussian-distributed random function, we can decompose the four-point correlation function $\langle S_{cc^\prime}(E)S^{\dagger}_{cc^{\prime}}(E)S_{cc^\prime}(E+\varepsilon)S^{\dagger}_{cc^{\prime}}(E + \varepsilon)\rangle$ into two terms:
\begin{eqnarray}
\langle S_{cc^\prime}(E)S^{\dagger}_{cc^{\prime}}(E)S_{cc^\prime}(E+\varepsilon)S^{\dagger}_{cc^{\prime}}(E + \varepsilon)\rangle && \nonumber \\
=
 \langle S_{cc^\prime}(E)S^{\dagger}(E)_{cc^{\prime}}(E)\rangle  \langle S_{cc^{\prime}}(E+\varepsilon)S^{\dagger}_{cc^{\prime}}(E+ \varepsilon)\rangle \nonumber \\
 \;\;\;\;+   
 \langle S_{cc^\prime}(E)S^{\dagger}_{cc^{\prime}}(E + \varepsilon)\rangle  \langle S^{\dagger}_{cc^{\prime}}(E)S_{cc^\prime}(E+\varepsilon)\rangle
\end{eqnarray}
The first term on the right-hand side is just the product of the average cross sections and the second term can be written as $|\langle S_{cc^\prime}(E)S^{\dagger}_{cc^{\prime}}(E + \varepsilon)\rangle|^2$. Thus, we have to calculate the $S$-matrix correlation function
\begin{equation}
C^{S}_{cc^{\prime}} = \langle S_{cc^\prime}(E)S^{\dagger}_{cc^{\prime}}(E + \varepsilon)\rangle \; .
\end{equation}

The calculation of the $S$-matrix correlation function follows exactly the same steps as that of the average cross section. The energy average involved is of the form
\begin{equation}
C^{S}_{cc^{\prime}} = \left<\sigma_{cc^{\prime}}(E)\right>  \langle \frac{1}{1 - i\varepsilon/\Gamma_q}\rangle_{q} =  \left<\sigma_{cc^{\prime}}(E)\right> \frac{1}{1 - i\varepsilon/\Gamma_{corr}},
\end{equation}
where $\Gamma_{corr}$ is called the correlation width. It is NOT equal to the average CN width $\Gamma$, but can be related to the sum of the transmission coefficients, $\frac{2\pi\Gamma_{corr}}{D} = \sum_{c} P_{cc}$. No such simple relation exists for the average width. 

The cross section correlation function can now be evaluated. Since $\langle\sigma_{cc^{\prime}}(E)\rangle =\langle\sigma_{cc^{\prime}}(E + \varepsilon)\rangle$, we have

\begin{equation}
C_(cc^{\prime})(\varepsilon) = \frac{(\langle\sigma_{cc^{\prime}}\rangle)^2}{1 + (\frac{\varepsilon}{\Gamma_{corr}})^2} \; ,
\end{equation}
where the average cross section is given by the Hauser-Feshbach expression, $\sigma^{CN}_{cc^{\prime}}$. In the presence of direct reactions the above Lorentzian form is maintained, but the numerator changes to
$2\sigma^{dir}_{cc^{\prime}} \sigma^{CN}_{cc^{\prime}} +  (\sigma^{CN}_{cc^{\prime}} )^2$. 

An Ericson analysis of cross section data supplies the correlation width $\Gamma_{corr}$. It has been verified experimentally that the experimental $C_{cc^{\prime}}(\varepsilon)$, for small enough $\varepsilon$, is a Lorentzian. This gave this method acceptance not only in nuclear physics research but also in other branches of physics.

%--Section: Preequilibrium reactions
\section{Preequilibrium reactions}
\label{sec_preeq}

The recognition that the spectra of particles in a given nuclear reaction invariably show deviations from pure compound nucleus emission or a fast direct process, led J. Griffin~\cite{Griffin-66} to propose the exciton model of pre-equilibrium emission. These reactions have attracted the attention of physicists ever since and a number of other semiclassical and quantum mechanical models of this emission process have been developed. Here we briefly describe the principal theoretical descriptions of pre-equilibrium emission and discuss challenges and new ideas in the area.

\subsection{Semiclassical models}
\label{subsec_preeq_semiclassical}

Griffin proposed the exciton model of preequilibrium emission~\cite{Griffin-66} to explain
an excess of high energy neutrons relative to those from the compound nucleus observed
in (p,n) reactions~\cite{Holbrow-63,Wood-65,Borchers-66}. 
In a proton-induced reaction, a collision of the proton with the nucleus leads to 
a 2p-1h state, in which the incident particle has
excited a target particles, creating a particle-hole pair.
At energies below about 200 MeV, the wavelength of a nucleon
is still greater than about 2 fm, so that the projectile nucleon - target
nucleus interaction actually does not excite an individual particle-hole state,
but some linear combination of these. Further
interaction of any of the particles or holes could create, scatter
or destroy particle-hole pairs. The exciton model classifies
the states in terms of the number of particles \( p \) and number
of holes \( h \) and assumes that all states with the same number of particles and holes are equally
populated.  Since the total number of particles is conserved, the difference between
the two, \( p-h \), remains constant throughout a collision. The quantity \( n=p+h \), is
called the exciton number. Particles emitted from configurations with low exciton number tend
to be more energetic and more focussed in the forward direction than those emitted from the
compound nucleus.

The competition between transitions to configurations of different exciton number
and emission is essential to the exciton model. Cline and Blann cast the model in the form of a
time-dependent master equation~\cite{Cline-71}, in which energy is conserved but angular momentum is not.
The equation governing the time development
of the fraction of the cross section \( P(n) \) in the \( n \) exciton
configuration is written as
\begin{eqnarray}
\label{preeqeq}
\frac{dP(n)}{dt}&=&\lambda _{-}(n+2)\, P(n+2)+\lambda _{0}(n)\, P(n) \nonumber \\
&& +\lambda _{+}(n-2)\, P(n-2)-\lambda (n)\, P(n)
\end{eqnarray}
 where \( \lambda (n) \) is the total rate of transitions out of
the \( n \) exciton configuration,\[
\lambda (n)=\lambda _{-}(n)+\lambda _{0}(n)+\lambda _{+}(n)+\lambda _{e}(n),\]
 with \( \lambda _{e}(n) \) being the total rate of particle emission
from the \( n \) exciton configuration. The quantities \( \lambda _{-}(n) \),
\( \lambda _{0}(n) \), and \( \lambda _{+}(n) \) are the average
rates for internal transitions from the \( n \) exciton configuration
with a change of exciton number by -2, 0, or +2. The average rate
of transitions that do not change the number of excitons, \( \lambda _{0}(n) \), cancels here.

Using Fermi's golden rule, the internal transition rates can, in principal,
be calculated by summing over all squared residual interaction matrix
elements leading from the initial to the final configuration. In practice,
this sum is written as the product of the average squared matrix element
of the residual interaction \( |M|^{2} \)  with the density of available
states. For the exciton-number-changing transitions,
the density of available states counts the average number of ways
an exciton of the initial particle-hole configuration
can be converted to three excitons (or vice versa), assuming the energy-conserving
transition between any of the single exciton states and any of the
three-exciton states to occur with equal likelihood. Similarly, the
density of available states for transitions that do not change the
exciton number counts the average number of ways any two
excitons may scatter from one another, again assuming equal likelihood
for all energy-conserving transitions. Expressions for the density
of available states were given by Williams~\cite{Williams-70} and later
corrected for the Pauli principle by Cline~\cite{Cline-72b}. The transition rate
$\lambda_{+}(n)$ can also be expressed in terms of the imaginary part of
the optical potential~\cite{Gadioli-73} or in terms of the in-medium nucleon-nucleon scattering
cross section or mean free path~\cite{Blann-71}. The particle-emission rate is written in
terms of the Weisskopf-Ewing emission rate using
the appropriate preequilibrium densities of states~\cite{Williams-71}.

The exciton model uses integrated transition and emission rates to determine the 
competition between scattering and emission.  The hybrid model~\cite{Blann-71,Blann-72,Blann-73}  
uses exciton model densities to determine the probability that a particle or hole has
a given energy, but then uses transition and emission rates for that energy to determine
the competition between escape and scattering. Results similar to those of the exciton model 
can be obtained by slightly modifying the overall magnitude of the transition rate.

Fundamental to both the exciton and the hybrid model is the assumption of equal occupation of the
states in each n-exciton configuration. Blann and Vonach demonstrated that this assumption is
fairly well satisfied by the initial transitions to the 2p-1h exciton configuration excited in a
nucleon-induced reaction~\cite{Blann-83}. Bisplinghoff showed, however, that this is not the case for 
the transitions to more complex configurations~\cite{Bisplinghoff-86}. The assumption would still
be reasonable 
if the transition rate among states of the same configuration, $\lambda_{0}(n)$, were much larger
than the configuration changing transition rates, $\lambda_{\pm}(n)$, so that
equilibration could occur before a subsequent transition. This is almost never the case, with
\begin{equation}
\lambda_{+}(n)>\lambda_{0}(n)>\lambda_{-}(n)
\end{equation}
for small $n$ for almost all excitation energies. This is due to the fact that the density of states
of the $n$ exciton configuration is relatively small compared to that of the $n+2$ configuration when $n$ is small. 

\begin{figure}[h!] 
\begin{center} 
    \includegraphics[scale=0.35]{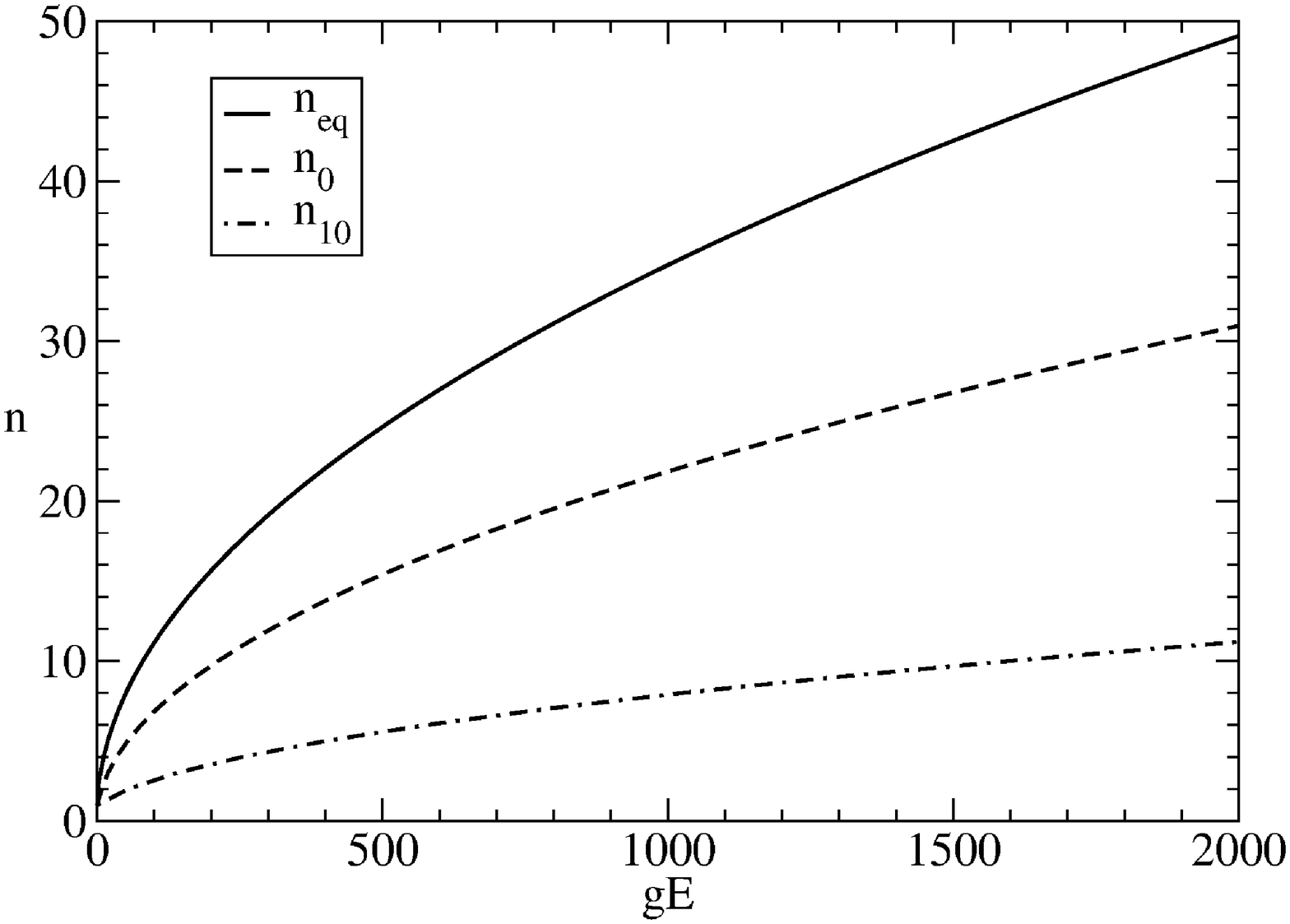}
     \caption{\it Average exciton number at equilibrium, $n_{eq}$, and minimum exciton number for equilibrium
within a configuration, $n_{0}$, as a function of $gE$, the product of the single particle state density and
the excitation energy. The additional curve, labeled $n_{10}$, is explained in the text.}
      \label{exciton-en}
\end{center}
\end{figure}

In Fig.\ref{exciton-en}, we show the exciton number of the
configuration defining complete equilibration, $n_{eq}$, for which    
\begin{equation}
\lambda_{+}(n_{eq})=\lambda_{-}(n_{eq}),
\end{equation}
as well as the minimum exciton number for equilbration within a configuration, $n_{0}$, which we define by
\begin{equation}
\lambda_{+}(n_{0})=\lambda_{0}(n_{0}),
\end{equation} 
as a function of the product of the single-particle density of states, $g\approx A/13$ MeV$^{-1}$ and the excition energy $E$.
We have assumed that the average matrix element inducing configuration changing transitions is 
equal to that among states of the same configuration. For reference, we also show the exciton number for 
equilibration within a configuration when the average squared matrix element for transitions among the states
of a configuration is 10 times that of the transitions between configurations, $n_{10}$. 
This can be the case at low excitation energy~\cite{Sato-87},
where hole-hole transitions dominate the $\lambda_{0}(n)$ transition rate.
However, as the excitation energy increases, the particle-hole and particle-particle matrix elements rapidly 
gain dominance in the transition rate, resulting in average matrix elements that are nearly equal.

Returning now to the question of equilibration among the states of a given configuration,
from Fig.\ref{exciton-en} we see that this is plausible for
the 3p-2h exciton configuration at values of $gE$ below about 50, corresponding to an excitation energy of about
17 MeV in $^{40}$Ca and of about 3 MeV in $^{208}$Pb. The minimum hole number increases as 
$n_{0}\approx\sqrt{0.4gE}$ while the equilibrium hole number increases as $n_{eq}\approx\sqrt{gE}$. Thus,
although the exciton and hybrid model do provide a reasonably good description of preequilibrium spectra, one of
their basic tenets is not satisfied. In part, the good agreement is due to the fact that at lower excitation energies, 
most emission occurs from the initial 2p-1h configuration. As the excitation energy increases and more complex
configurations contribute to emission rates, discrepancies would be expected.

To avoid this problem, Blann proposed the hybrid
Monte Carlo simulation model~\cite{Blann-96}, which he later extended, in collaboration with Chadwick, to a model of
double differential preequilibrium spectra~\cite{Blann-98a}. Like the hybrid model, it calculates transition rates
using the energy of each particle or hole. However, it only uses the transition rates to 2p-1h and 1p-2h configurations
corresponding to the individual particles and holes and determines the energy of each particle and hole after each collision, using
Monte Carlo selection from the 2p-1h and 1p-1h distributions. In a sense, the model proposes an intranuclear cascade that is 
performed in energy-angle space rather than configuration space. The model is more consistent than the exciton 
and hybrid models and can describe pre-equilibrium spectra and double differential data at least as well as these.
However, like these, it generally does not describe well the high energy component of spectra, where collective effects are
important, nor scattering at back angles, where quantum coherence effects play a role.

\subsection{Quantum models}  

A fundamental difference between semiclassical and quantum models of
preequilibrium reactions is that semiclassical models do not distinguish between 
bound and continuum states, while quantum models must do so.
The exciton picture is still useful here, as a means to refine the Feshbach decomposition of the
space of states into a continuum component \( P \) and a bound state
component \( Q \). Maintaining our emphasis on nucleon-induced reactions, we can write
\begin{eqnarray}
P & =P_{1}+ & P_{3}+P_{5}+P_{7}+...\, ,\nonumber\\
Q & =\quad \; \; \;  & Q_{3}+Q_{5}+Q_{7}+...\, .\label{pnsnqns} 
\end{eqnarray}
 The decomposition of the direct reaction space \( P \) contains
the elastic component \( P_{1} \) and a series of components with
increasing exciton number in which it is assumed that one (and only
one) of the nucleons is in the continuum. Progression along
the stages of the \( P \) chain is usually assumed to result from additional
interactions of the continuum nucleon with the target, although it could also
occur due to interactions
within the target. The decomposition of the compound nucleus space
begins with the three-exciton configuration \( Q_{3} \), since the
incident nucleon must collide with at least one nucleon, forming a
two-particle, one-hole configuration, to be captured into
a quasi-bound state. Transitions between the chains may also occur
at any stage. However, it is assumed that all transitions change the
exciton number by at most two.

Reactions that occur in the direct reaction space \( P \) are known
as multistep direct reactions. Those that occur in the compound nuclear
space \( Q \) are known as multistep compound reactions. The first
multistep direct models were developed by Feshbach, Kerman and
Koonin~\cite{Feshbach-80}
and by Tamura, Udagawa and Lenske~\cite{Tamura-82} and later by Nishioka,
Weidenm\"{u}ller, and Yoshida~\cite{Nishioka-88}. The first multistep compound
model was developed by Agassi, Weidenm\"{u}ller and Mantzouranis~\cite{Agassi-75}
and rederived using more rigorous methods in Ref. ~\cite{Nishioka-86}.
Similar models were also proposed in Refs.~\cite{Feshbach-80,Friedman-81,McVoy-83}.

\subsubsection{The multistep compound model}

The multistep compound model
includes in its state or level densities only those
states in which all of the single-particle states are bound.
At first glance, it would thus appear to be an exciton model in which
the transitions are limited to only bound single-particle 
states~\cite{Stankiewicz-85,Oblozinsky-86}.
In distinction from the exciton model, the multistep
compound one requires an interaction to occur for
a nucleon to be emitted from the composite system. Absorption occurs
much as in the exciton model, although the transition must lower the initially
unbound nucleon to a bound configuration. In
the case of emission, the transition must raise one of the
nucleons to an unbound, continuum state so that it can leave the system.
Like the internal transition factors, $\lambda_{\pm}(n)$ and $\lambda_{0}(n)$,
the factors describing emission can change the
exciton number by two, $ Y_{c\pm}(n)$ or leave it unchanged  $Y_{c0}(n)$
(taking into account the particle in the continuum).
However, one-particle to two-particle, one-hole transitions
cannot contribute to the emission rate, due to the restriction to
bound initial one-particle configurations. 

The absorption factors describe
the inverse process to emission and may be obtained by reinterpreting the
emission factors. In particular, the absorption factor for creating the initial 
\( 2p-1h \) configuration is \( Y_{c-}(3) \). This form of the
multistep compound emission/absorption factors was first derived
in Ref.~\cite{Feshbach-80}. Explicit expressions for these and a general
discussion of multistep compound processes can be found in Ref.~\cite{Bonetti-91}.

A nucleon that undergoes a transition to the continuum can return
to a bound-state configuration due to another transition before escaping
the nucleus. This introduces a component proportional
to the product of an emission and an absorption factor, called
the external mixing component. It permits transitions (through the continuum)
that change the exciton number by up to 4. It is derived and discussed
in Ref. ~\cite{Agassi-75}, but, to our knowledge, has not been included in multistep
compound calculations.

Because the multistep compound model requires that all particles be in bound states,
a multistep compound reaction is initiated from a multistep direct stage at all but the lowest energies.
Depending on the initial energy, the multistep compound process can be fed by
early, intermediate or late stages of the multistep direct chain. As such
transitions initiate a multistep compound chain at a higher exciton
number than those that enter it directly from the nuclear ground state,
they tend to decrease multistep compound preequilibrium emission
rather than increase it. Most preequilibrium multistep compound emissions
come from the first few stages in the chain. A composite system formed
with a higher exciton number has a greater chance of evolving to equilibrium
before decaying than one formed with a smaller exciton number.

At low to medium excitation energy, the multistep compound transition rates of low exciton number retain the
same proportionality as those of the exciton model~\cite{Stankiewicz-85,Oblozinsky-86}.
The multistep compound model thus suffers from the same conceptual difficulty as the exciton and hybrid models:
the assumption of equal occupation of states in configurations with more than three excitons cannot be
justified. This difficulty does not necessarily diminish as the energy increases and a multistep direct stage
becomes necessary, since the minimum exciton number for equilibration between transitions also increases with energy.

\subsubsection{The multistep direct model}

To obtain expressions for multistep direct reactions, one
analyzes the coupling along the chain of continuum configurations 
\begin{equation}
P =P_{1}+ P_{3}+P_{5}+P_{7}+...
\end{equation}
and, of course, their coupling to the compound nucleus states. 
One assumes that the Hamiltonian can be written as 
\begin{equation}
H=H_{0}+V
\end{equation}
where $H_{0}$ consists of the projectile and target Hamiltonian, as well as
a projectile-target optical potential that accounts for the flux lost due to
the interaction $V$ but does not induce transitions. All transitions are
assumed to be the result of the interaction $V$. The excitations
of the multistep direct model are assumed to be individually
weak, although large in number, so that it is advantageous to rewrite
the Schr\"{o}dinger equation as a Lippmann-Schwinger equation,
\begin{equation}
(E-H_{0}-V)\left|\psi^{(+)}\right>=0\quad\longrightarrow\quad\left|\psi^{(+)}\right>=\left|\phi^{(+)}\right>+\frac{1}{E-H_{0}}V\left|\psi^{(+)}\right>,
\end{equation}
 where \( (E-H_{0})\left|\phi^{(+)} \right>=0, \) and to approximate the wavefunction
using a series expansion,
\begin{equation}
\left|\psi^{(+)}\right> = \sum_{n=0}\left(\frac{1}{E-H_{0}}V\right)^n \left|\phi^{(+)}\right>\,.
\end{equation}
The scattering amplitude can then be written as 
\begin{eqnarray}
T_{\mu0}  &=&  \left<\phi^{(-)}_{\mu}\right|V\left|\phi^{(+)}_{0}\right> +
\left<\phi^{(-)}_{\mu}\right|V\frac{1}{E-H_{0}}V\left|\phi^{(+)}_{0}\right> \nonumber \\
&& +\left<\phi^{(-)}_{\mu}\right|V\frac{1}{E-H_{0}}V\frac{1}{E-H_{0}}V\left|\phi^{(+)}_{0}\right> +\dots
\end{eqnarray}
To reduce the cross section to a sum of terms, one observes that the first term projects onto the
1p-1h component of the final states, the second onto the 2p-2h component, and so forth (assuming
of course that no transitions are induced by the target Hamiltonian in $H_{0}$). One then argues that
averaging over a small interval in final excitation energy renders the contributions from different
configurations incoherent relative to one another, reducing the cross section to a sum of terms,
\begin{eqnarray}
\frac{d^2\sigma}{dE_md\Omega} & = & \overline{\left|\left<\phi^{(-)}_{m}\right|V\left|\phi^{(+)}_{0}\right>\right|^2}+\overline{ \left|\left<\phi^{(-)}_{m}\right|V\frac{1}{E-H_{0}}V\left|\phi^{(+)}_{0}\right>\right|^2} \\
& & \qquad +\overline{ \left|\left<\phi^{(-)}_{m}\right|V\frac{1}{E-H_{0}}V\frac{1}{E-H_{0}}V\left|\phi^{(+)}_{0}\right>\right|^2} +\dots\nonumber
\end{eqnarray}
Other statistical hypotheses are necessary to simplify the expression further.

Multistep compound models have been carried out to various orders and with 
varying degrees of sophistication. In the model proposed by Feshbach, Kerman
and Koonin, an average residual interaction is generally used with exciton configuration
densities. The statistical hypotheses are assumed to be sufficiently strong to
reduce the n-step cross section to a convolution of single step cross sections.
A two-step cross section, for example, could then be written as
\begin{equation}
\frac{d^2\sigma}{dEd\Omega}\left(\vec{k},\vec{k}_0\right)=\int\frac{d^3k^{\prime}}{2\pi^3}\,\frac{d^2\sigma}{dEd\Omega}\left(\vec{k},\vec{k}^{\prime}\right)\,\frac{d^2\sigma}{dEd\Omega}\left(\vec{k}^{\prime},\vec{k}_0\right)
\end{equation}
The resulting expressions are intuitively pleasing and fairly easy to calculate.
Because of this, the model has been compared extensively to experimental data~\cite{Bonetti-94}
and used to calculate up to fourth order and charge-exchange reactions~\cite{Koning-97}. However,
reduction of the n-step cross sections to convolutions requires the substitution of
an optical wave function with an outgoing boundary condition by one with an incoming boundary condition,
\(\phi^{(+)}\rightarrow \phi^{(-)}\), in the intermediate propagators. As these wavefunctions differ 
in magnitude by a factor of the optical S matrix, the approximation
has proven difficult to justify.

In an alternative approach, Tamura, Udagwa and Lenske~\cite{Tamura-82,Ramstrom-04} 
developed the physical structure of the excitations in more detail, but limited
the reaction chain to two steps. To obtain the one-step cross section, they 
calculated a 1p-1h response function by diagonalizing the target nucleus
Hamiltonian in the 1p-1h subspace. The advantage of such an approach is
that it can take into account both the collective and the single-particle response
of the excitation. Such single-step calculations have been performed with even more rigor 
in recent years and show excellent agreement with experimental
data~\cite{Dupuis-06a,Dupuis-11}. 

As the incident energy increases, further direct reaction stages become necessary. Tamura,
Udagawa and Lenske extended their one-step response function to a two-step one by assuming
that the microscopic structure of the intermediate states in the two-step cross section
is incoherent  on the average. They then convoluted the 1p-1h response with itself to
obtain the 2p-2h response function, but retained the intermediate propagator form of the spatial structure of the 
two-step cross section. Nishioka, Weidenm\"uller and Yoshida claim that the
statistics is insufficent even for averaging the microscopic structure and that the
coherent 2p-2h response function should be used in the two-step cross section. We build on arguments
they put forth in Ref.~\cite{Nishioka-88} to analyze the assumptions commonly used in multistep direct models. 
\begin{figure}[h!] 
\begin{center} 
    \includegraphics[scale=0.35]{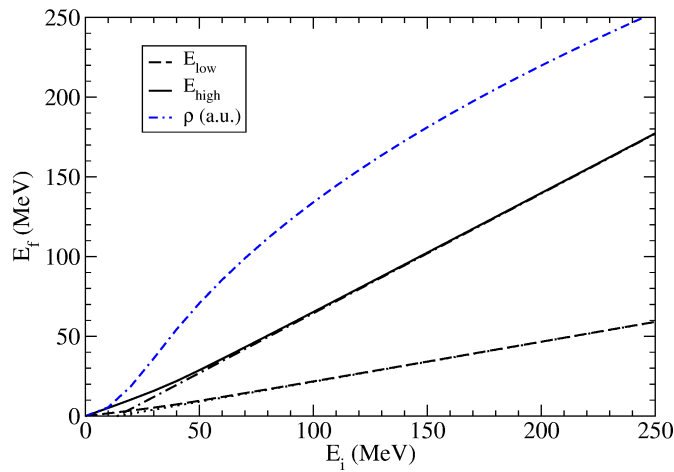}
     \caption{\it Average energy of the fast (leading) and slow final particles in a 
nucleon-nucleon collision in nuclear matter as a function of the incident particle energy. Also
shown is the dependence on initial energy of the final phase space volume for a collision, in arbitrary units.}
      \label{ave-en}
\end{center}
\end{figure}

We first analyze the approximations inherent in the 'leading particle' nature of the model,
that is, in the fact that only one nucleon is assumed to be in the continuum and is followed 
through the reaction. In Fig.\ref{ave-en}, the average energy of the fast (leading) and slow
final particles in a nucleon-nucleon collision (assumed to be isotropic in the center-of-mass)
in nuclear matter are displayed
as a function of the incident particle energy~\cite{Kikuchi-68}. Above about 50 MeV, the
averages values of the two energies are well approximated by the straight lines in the figure,
\begin{equation}
E_{high}=\frac{3}{4}\left(E_i-\left<E_{hole}\right>\right) \;\;\; \mbox{and} \;\;\;
E_{low}=\frac{1}{4}\left(E_i-\left<E_{hole}\right>\right)
\end{equation}
where the average hole energy is  \(\left<E_{hole}\right> = 2E_{F}/7\), $E_F$ being the Fermi energy of
the (nonrelativistic) Fermi gas. The widths of the energy distributions of the fast and
slow particles are both about \(E_i/7\). Also shown in the figure is the incident energy
dependence of the final phase space volume for the collision, in arbitrary units, which we take as
a relative measure of the probability of a collision. 

If we consider an incident particle of 60 MeV, we conclude that
the  energy of the leading particle after the collision will be about 30$\pm$9 Mev and the 
that of the slow particle about 10$\pm$9 MeV, so that both are probably unbound. If we
compare the final phase space volumes 
associated with the two, we find that a second collision is about 7 times more likely to be
induced by the 30 MeV particle than by the 10 MeV one. Collision of the 30 MeV particle
with another nucleon in the Fermi sea, would produce a leading particle of 12$\pm$4 MeV, still unbound, 
on the average. 

This problem becomes more serious as the energy increases. If we consider a 150 MeV incident nucleon,
we find that the average energy of the fast particle after the collison will be about 100$\pm$20 MeV,
while the average energy of the slow one will be about 35$\pm$20 MeV. In this case the probability of
a second collision being induced by the fast particle is only about 3 times greater than that of being
induced by the slow particle. The leading particle model of multistep direct reactions is thus
limited to reactions at fairly low incident energies for two reasons. First, at energies above about 30 MeV,
the probability of the slower particle in a collison being unbound becomes appreciable. Second, as the energy
increases, so does the probability that the slow particle interacts before the leading particle
does so again.

However, even at low incident energies, the multistep direct model suffers from the same problem as the exciton
and multistep compound models: for low exciton number, the transition rate increasing the exciton number is much larger
than the transition rate mixing states with the same exciton number. Here the lack of equilibration among the states
of same exciton number implies that they cannot be considered to contribute incoherently, as it is the interaction 
among these states that would turn their phases random. A correct description of the two-step process
would thus require the partially coherent 2p-2h response function. At higher energies,
where the three step process should be included, one would also need the complete 3p-3h response function,
since the limiting exciton number for equilbrium before transition, $n_{0}$, increases with energy as well. 

 We conclude from the above analysis that, with the exception of reactions at very low energies
in which the compound or multistep compound nucleus is formed directly, the chain of preequilibrium
reactions that eventually lead to the compound nucleus are direct reactions and are at least partially
coherent over a wide range of exciton numbers. In addition, at energies greater than about 30 MeV, these 
reactions will result in more than one continuum particle, as well as particles in bound states, that 
can induce further transitions. Neither the effects of coherence beyond that in 1p-1h excitations
nor the contributions of transitions induced by particles other than the leading
one have been included in multistep direct reactions. The latter are contained in the semiclassical
hybrid Monte Carlo simulation model proposed by Blann and Chadwick~\cite{Blann-98a}. However, its semiclassical
nature inhibits inclusion of the collective and diffractive effects of quantum scattering that are
essential to a complete description of preequilbrium reactions.  

Before closing, we observe that pickup reactions leading to composite particles such as the deuteron,
tritium and alpha are also important in preequilibrium scattering and can contribute up to 20\% of the
preequilibrium yield~\cite{Bertrand-73}. Preequilibrium emission occurs in reactions induced by
composite particles as well, which can produce both pickup and stripping components. In all cases, composite
particle emission can occur directly or from a mixture of continuum and 
bound states~\cite{Griffin-67,Blann-70,Iwamoto-82,Sato-83}. Although preequilibrium reactions have been studied
for many years now, we are still only beginning to learn how to describe them.

%--Section: Hybrid reactions
\section{Hybrid reactions}
\label{sec_hybrids}

Nuclear reaction cross sections are required input for models of stellar evolution and element synthesis, simulations of the nuclear fuel cycle, and other applications. Of interest are reactions of neutrons and light, charged particles with target nuclei across the isotopic chart, taking place at energies from several keV to tens of MeV.  Unfortunately, for a large number of reactions the relevant data cannot be directly measured in the laboratory or easily determined by calculations. 

Direct measurements may encounter a variety of obstacles:  The energy regime relevant for a particular application is often not accessible -- cross sections for charged-particle reactions, e.g., become vanishingly small as the relative energy of the colliding nuclei decreases.  Furthermore, many important reactions involve unstable nuclei which are too difficult to produce with currently available techniques, too short-lived to serve as targets in present-day set-ups, or highly radioactive.  Producing all relevant isotopes will remain challenging even for radioactive beam facilities.  

Cross section calculations require a thorough understanding of both direct and statistical reaction mechanisms (as well as their interplay) and a detailed knowledge of the nuclear structure involved. 
Nuclear-structure models can provide only limited information and very little is known about important quantities such as optical-model potentials or spectroscopic factors for nuclei outside the valley of stability.

In order to overcome these limitations, one has to resort to indirect methods, which normally rely on a combination of theory and experiment for success. 
Approaches such as the Asymptotic Normalization Coefficient (ANC) method~\cite{Xu:94,Mukhamedzhanov:01}, Coulomb dissociation~\cite{Baur:96,Baur:03}, and the Trojan-Horse method~\cite{Typel:03} have yielded valuable cross-section information for various direct reactions. 
Here we focus on the {\em Surrogate Nuclear Reactions technique}, which employs a transfer or inelastic scattering reaction to determine a compound-nucleus cross section indirectly.

\subsection{The Surrogate Method}
\label{sec_hybrids_surrogate}

The surrogate nuclear reaction approach combines experiment with theory to obtain cross sections for CN reactions, $a+A \rightarrow B^* \rightarrow c+C$, involving difficult-to-produce targets, $A$~\cite{Britt:70,Britt:79,EscherDietrich:06,Escher:10a,Escher:12rmp}. Of particular interest are neutron-induced reactions on unstable nuclei, such as neutron capture $A(n,\gamma)B$ (e.g. $^{95}Zr(n,\gamma)^{96}Zr$) and neutron-induced fission (e.g. $^{240}Am(n,f)$).
In the Hauser-Feshbach formalism~\cite{HauserFeshbach:52}, the cross section for this ``desired'' reaction takes the form:
\begin{eqnarray}
\sigma_{\alpha \chi}(E_{a}) &=& \sum_{J,\pi}  \sigma_{\alpha}^{CN}(E,J,\pi) \;\; G_{\chi}^{CN}(E,J,\pi) \; ,
\label{eq:DesReact}
\end {eqnarray}
\noindent
with $\alpha$ and $\chi$ denoting the relevant entrance and exit channels, $a+A$ and $c+C$, respectively.  The excitation energy $E$ of the compound nucleus, $B^*$, is related to the projectile energy $E_a$ via the energy needed for separating a from $B$: $E_a=E-S_a(B)$.  In many cases the formation cross sections $\sigma_{\alpha}^{CN}(E,J,\pi)$, which is directly related to the transmission coefficient $T_c$ in Eq.~\ref{eq:HF}, can be calculated to a reasonable accuracy by using optical potentials. On the other hand, the theoretical decay probabilities $G_{\chi}^{CN}(E,J,\pi)$ for the different decay channels (which are essentially the ratios $T_c$/$\sum_{c''}T_{c''}$ in the same expression, Eq.~\ref{eq:HF}) are often quite uncertain. The latter are difficult to calculate accurately since they require knowledge of optical models, level densities, and strength functions for the various possible exit channels.  The objective of the surrogate method is to determine or constrain these decay probabilities experimentally.

In a surrogate experiment, the compound nucleus $B^*$ is produced by means of an alternative (``surrogateÓ) reaction, $d+D \rightarrow b+B^*$ that utilizes a target-projectile combination that is experimentally more accessible. The surrogate reaction might involve inelastic scattering, such as $(p,p')$, pickup, or stripping reactions, such as (d,p) in inverse-kinematics setups. In the experiment, the desired decay channel $\chi (B^* \rightarrow c+C$) is observed in coincidence with the outgoing particle $b$.  The coincidence measurement provides
\begin{eqnarray}
P_{\delta\chi}(E) &=& \sum_{J,\pi} F_{\delta}^{CN}(E,J,\pi) \;\; G_{\chi}^{CN}(E,J,\pi) \; ,
\label{eq:SurReact}
\end {eqnarray}
the probability that the compound nucleus was formed in the surrogate reaction with spin-parity distribution $F_{\delta}^{CN}(E,J,\pi)$ and subsequently decayed into the channel $\chi$.  
The spin-parity distributions $F_{\delta}^{CN}(E,J,\pi)$, which may be very different from the compound-nuclear spin-parity populations following the absorption of the projectile $a$ in the desired reaction, have to be determined theoretically, so that the branching ratios $G_{\chi}^{CN}(E,J,\pi)$ can be extracted from the measurements.  
In practice, the decay of the compound nucleus is modeled and the $G_{\chi}^{CN}(E,J,\pi)$ are obtained by fitting the calculations to reproduce the measured decay probabilities.  Subsequently, the sought-after cross section can be obtained by combining the calculated cross section $\sigma_{\alpha}^{CN}(E_{ex},J,\pi)$ for the formation of $B^*$ (from $a+A$) with the extracted decay probabilities $G_{\chi}^{CN}(E_{ex},J,\pi)$ for this state (see Eq.~\ref{eq:DesReact}).

Predicting the spin-parity distribution for a \cn produced in a surrogate reaction requires a careful consideration of the reaction mechanisms that are involved in the formation of the \cnn.  In the absence of width fluctuation corrections, the challenge of describing the relevant reaction mechanisms can be divided into two separate problems:
\begin{itemize}
\item[1)] the formation of a highly-excited nucleus in a direct reaction, and 
\item[2)] the damping of the excited states into the compound nucleus.  
\end{itemize}
The separation of the surrogate reaction into two separate sub-processes is somewhat artificial, but may be useful conceptually.  The surrogate reaction is viewed as a mechanism that produces initially a highly-excited intermediate system.  The system might consist, for instance, of a nucleon $N$ (stripped from the projectile $d$ in the reaction $d+D \rightarrow b+B^*$) plus the surrogate target nucleus $D$.  For the surrogate approach to be valid, the $D+N$ system must subsequently fuse to produce the compound nucleus $B^*$, the decay of which one is interested in measuring.
Decay of the intermediate system ($D+N$ in the example) by particle emission prior to reaching the equilibrated stage would invalidate the surrogate approach, since the measured coincidence probabilities would no longer be associated with the decay of the \cn of interest, $B^*$.  It is thus important to understand how the configurations that are produced in the initial step evolve.  Specifically, one needs to determine the probability for forming the desired \cn $B^*$.

Addressing the first problem necessitates developing a quantitative description of the direct-reaction process that allows for a prediction of the spin-parity distribution in the highly-excited intermediate nucleus, immediately following the direct reaction. Such a description requires a framework for calculating cross sections of different reactions (stripping, pick-up, charge exchange, and inelastic scattering) to continuum states, for a variety of projectiles ($p$, $d$, $t$, $\alpha$, etc.) and targets (spherical, deformed, and transitional). 
First steps towards predicting the spin-parity population following the initial step of a surrogate reaction were taken by Andersen \etal \cite{Andersen:70}, Back \etal \cite{Back:74a}, and Younes and Britt~\cite{Younes:03a,Younes:03b}.  These authors employed simple transfer calculations to estimate \cndash spin-parity distributions following various stripping reactions on actinide targets. They neglected the possibility that the intermediate nucleus might decay prior to reaching equilibrium and took the resulting spin-parity distributions to be representative of those present in the \cn created in the surrogate reaction of interest.  Modeling the decay of the compound nuclei allowed them to extract (n,f) cross sections which showed good agreement with directly measured cross sections, where those benchmarks existed.
 
The second problem to be addressed is associated with the evolution of the highly-excited intermediate system that is created in the initial stage of the surrogate reaction. The assumption that a compound (i.e. equilibrated) nucleus is formed is central to the surrogate method. Rapid decay of the intermediate configuration before a compound nucleus can be formed needs to be excluded experimentally, or accounted for theoretically. Recent studies of inelastic $^{3}$He scattering on Zr and Y targets, for instance, indicate that there is a small but non-negligible (5-20\%) contribution from `non-equilibrium' decay when excitation energies above 15 MeV or higher are reached in the intermediate nucleus~\cite{Escher:13cnr}. Discrepancies are visible between measured $\gamma$-ray intensities from the decay of the intermediate nuclear system and predictions from Hauser-Feshbach-type calculations that account for the surrogate spin-parity distribution but assume a fully equilibrated system.
Non-equilibrium decay is expected to play an even larger role in stripping reactions with weakly-bound projectiles. Therefore, it will affect the interpretation of inverse-kinematics $(d,p)$ experiments with radioactive beams.  

For obtaining a better understanding of the evolution of the intermediate, highly-excited, system following a direct reaction, ideas from the KKM approach (Section~\ref{sec_cnTheory_KKM} and Refs.~\cite{Kawai:73,Kerman:79}) can be employed. Some of these ideas have been investigated by F.S. Dietrich~\cite{Dietrich:07cnr} and by Parker \etal~\cite{Parker:95shrt} for an analogous case involving radiative neutron and proton capture.

%--Section: Challenges
\section{Some challenges and open questions}
\label{sec_Challenges}

The description of CN reactions requires that many detailed features of nuclear structure be ignored and the reaction be treated in some average sense. In the theories discussed here, this is accomplished by projecting out closed reaction channels and introducing energy averages. 
The underlying assumptions are that the reaction excites a sufficient number of states in the compound nucleus and that the direct-reaction contributions to the cross section do not vary significantly over the energy range considered. These conditions are typically satisfied when the beam energy spread is much larger than the CN level spacing and, at the same time, small compared to the incident energy.
It has long been known that these conditions are not always fulfilled, e.g. for reactions on light nuclei, when the density of CN resonances becomes too small for a statistical treatment, but the number of resonances remains too large to allow for an individual treatment~\cite{Hodgeson:book}.
Similar difficulties can be expected for the application of reaction theories to nuclei away from the valley of stability. Here the level densities decrease and inverse-kinematics experiments at rare isotope accelerators will have a limited range of beam energies (per nucleon) available.

Due to beam constraints, experiments at rare isotope facilities will encounter reactions that contain contributions from both compound and direct reactions.   To disentangle the components, it becomes important to describe them in a coherent framework. The KKM approach discussed in Section~\ref{sec_cnTheory_KKM} makes a step into that direction, by providing an extension of the Hauser-Feshbach formalism to the case where direct reactions are present.
The KKM theory is valid in the limit of strong absorption, defined by the condition that the CN resonance widths are large compared to their spacings, $\Gamma/D \gg 1$. 
Compound-nuclear cross section for the more general case, including weak absorption, $\Gamma/D \ll 1$, can in principle be calculated in the  framework of Random Matrix Theory, under the assumption of a Gaussian Orthogonal Ensemble (GOE) of real and symmetric Hamiltonian matrices~\cite{Mitchell:10}. This method uses ensemble averages, which are considered equivalent to the energy averages considered in this paper. The GOA approach has been very successful, both in the context of nuclear physics and for the description of other chaotic systems. However, it requires a computationally expensive calculation of a triple integral that enters the closed-form expression for the average cross section~\cite{Verbaarschot:85}.  

While the GOE approach is often considered a reference to which other theories should be compared, approximate methods are needed for practical applications. For instance, correlations between the different reaction channels are known to play an important role in certain circumstances (see Sections~\ref{sec_cnTheory_Bohr} and \ref{sec_cnTheory_KKM}). Generally, the corrections increase the compound-elastic cross sections, while reducing the cross sections for the other channels. The corrections are most important in the vicinity of thresholds and when few channels are open. Modern reaction codes account for these correlations via a width fluctuation corrections factor $W_{cc^{\prime}}$ (eq.~(\ref{eq:WFCF})).
In recent years, there has been renewed focus on testing various approximate expressions for $W_{cc^{\prime}}$ against the GOE approach~\cite{Hilaire:03} and to develop improvements~\cite{Kawano:ND2013}. Some of the assumptions underlying the KKM theory discussed here have also been tested against GOE predictions~\cite{Arbanas:08cnr,Arbanas:12cnr}.  In addition, new procedures for gaining insights into the correlations within the Random Matrix approach are currently considered~\cite{Ericson:13}. 

Pre-equilibrium processes give important contributions to many measured reaction cross sections. The underlying reaction mechanism involves a series of relatively weak interactions leading to an extremely large number of final states, so that the usual direct-reaction or coupled channel theory no longer applies. At the same time, a statistical compound nucleus type of description cannot be used either, as equilibrium is not reached. Various models of pre-equilibrium reactions have been developed. Modern reaction codes typically implement one or more of these. In Section~\ref{sec_preeq}, we discussed the most common approaches and outlined some of their short-comings. We concluded there that the existing preequilibrium models still fall far short of including the physics necessary for a general description of these reactions.

Understanding the interplay of direct and compound reaction mechanisms becomes very important for extracting cross sections indirectly from surrogate measurements, as well as for interpreting experiments with rare isotopes. 
In the context of surrogate reactions (see Section~\ref{sec_hybrids}), what is typically thought of as a direct-reaction process is used to initiate the formation of a compound nucleus.  The outgoing direct-reaction particle is detected in coincidence with reaction products that presumably result from the decay of the compound nucleus and conclusions are drawn about the decay characteristics of a related reaction. The accuracy of the extracted cross section depends on how well the compound-nucleus formation process in the surrogate reaction is described.
Similarly, separating the compound and direct-reaction contributions in low-energy radioactive beam experiments is crucial for obtaining reliable nuclear structure information.

%--Section: Summary
\section{Summary}
\label{sec_summary}

Over the decades, there have been extensive theory developments aimed at properly describing reactions involving compound nuclei. We have summarized here some important approaches that focus on the reaction mechanisms (rather than on the structure models that provide complementary input). Despite the wide use of the Hauser-Feshbach formalism in modern reaction codes, some of the available theoretical developments have not yet been implemented and some of the underlying assumptions remain to be tested.  Here we have revisited three areas of interest: the general theory of compound-nuclear reactions, the role of pre-equilibrium reactions, and extensions needed in order to get cross section information indirectly via hybrid reactions. We find that various questions and challenges remain to be addressed in order to gain a deeper understanding of the many-body processes involved and to obtain reliable, comprehensive descriptions of this important class of nuclear reactions.

\section*{Acknowledgments}
J.E.E. thanks F.S. Dietrich, D. Gogny, and A.K. Kerman for many insightful discussions on the subject of compound-nuclear reactions. Her work was performed under the auspices of the U.S. Department of Energy by Lawrence Livermore National Laboratory under contract DE-AC52-07NA27344, with partial support through the DOE's topical collaboration TORUS and the ASC/PEM program at LLNL.
B.V.C. acknowledges partial support from CAPES(Brazil), the CNPq(Brazil) under project 305574/2009-4 and FAPESP(Brazil) under project 2009/00069-5.
M.S.H. acknowledges partial support from the CNPq(Brazil) under project 311177/2010-7 and FAPESP(Brazil) under project 2011/18998-2.

%----- References -------------------------------

\end{document}